\documentclass[11pt]{article}

\usepackage{latexsym,times}
\usepackage{amsmath}
\usepackage{amsxtra}
\usepackage{amscd}
\usepackage{amsthm}
\usepackage{amsfonts}
\usepackage{amssymb}
\usepackage{dsfont}
\usepackage{xcolor}
\usepackage{cancel}
\usepackage{framed}

\usepackage{esint}

\usepackage{enumitem}
\usepackage{float}

\usepackage{hyperref}
\usepackage[square,sort&compress,comma,numbers]{natbib}

\makeatletter

\@addtoreset{equation}{section} \makeatother

\input amssym.def
\input amssym.tex

\newcommand\mr{\mathring}

\newcommand{\half}{{{\textstyle\frac{1}{2}}}}
\newcommand{\quarter}{{{\textstyle\frac{1}{4}}}}
\newcommand{\be}{\begin{equation}}
\newcommand{\ee}{\end{equation} }
\newcommand{\beqa}{\begin{eqnarray} }
\newcommand{\eeqa}{\end{eqnarray} }
\newcommand{\ba}{\begin{array}}
\newcommand{\ea}{\end{array}}
\newcommand{\bpm}{\begin{pmatrix}}
\newcommand{\epm}{\end{pmatrix}}

\newcommand{\Spin}{\mathbf{Spin}}

\newcommand{\Pin}{\mathbf{Pin}}

\newcommand{\deltaS}{\delta_{\varepsilon}}
\newcommand{\deltak}{\delta_{\kappa}}
\newcommand{\deltaCG}{\delta_{{\scriptscriptstyle{\rm{C.G.\,}}}}}
\newcommand{\deltaV}{\delta_{{\scriptscriptstyle{{\cV}}}}}

\newcommand{\ODD}{\mathbf{O}(D,D)}

\newcommand{\SpinD}{{\Spin(1,D{-1})}}
\newcommand{\oSpinD}{{{\Spin}(D{-1},1)}}

\newcommand{\Ott}{\mathbf{O}(10,10)}

\newcommand{\Spint}{{\Spin(1,9)}}
\newcommand{\oSpint}{{{\Spin}(9,1)}}

\newcommand{\oPint}{{{\Pin}(9,1)}}

\newcommand{\eleven}{{(11)}}
\newcommand{\gammae}{\gamma^{\eleven}}
\newcommand{\brgammae}{\brgamma^{\eleven}}

\newcommand\rd{{\rm d}}

\newcommand\rdA{D}

\newcommand\cA{{\cal A}}

\newcommand\cF{{\cal F}}

\newcommand\cH{{\cal H}}

\newcommand\cJ{{\cal J}}

\newcommand\cL{{\cal L}}

\newcommand\cS{{\cal S}}

\newcommand\cU{{\cal U}}
\newcommand\cV{{\cal V}}

\newcommand\hcL{{\hat{\cal L}}}

\newcommand\hepsilon{\hat{\epsilon}}

\newcommand\hlambda{\hat{\lambda}}
\newcommand\hmu{\hat{\mu}}
\newcommand\hnu{\hat{\nu}}
\newcommand\hrho{\hat{\rho}}
\newcommand\heta{\hat{\eta}}

\newcommand\zetap{{\zeta^{\prime}}{}}
\newcommand\thetap{\theta^{\prime}}
\newcommand\brthetap{\brtheta^{\prime}}

\newcommand\kappap{\kappa^{\prime}}

\newcommand\varepsilonp{\varepsilon^{\prime}{}}
\newcommand\brvarepsilon{\bar{\varepsilon}}
\newcommand\brvarepsilonp{\brvarepsilon^{\prime}{}}

\newcommand\alphap{{\alpha^{\prime}}}

\newcommand\h{{h}}

\newcommand\dis{\displaystyle}

\def\tx{\tilde{x}}

\def\tpartial{\tilde{\partial}}

\def\tY{\tilde{Y}}

\def\bre{\bar{e}}

\def\breta{\bar{\eta}}
\def\bralpha{\bar{\alpha}}
\def\brbeta{\bar{\beta}}
\def\brgamma{\bar{\gamma}}
\def\brdelta{\bar{\delta}}

\def\brp{{\bar{p}}}
\def\brq{{\bar{q}}}

\def\brC{\bar{C}}

\def\brV{{\bar{V}}}

\def\brP{{\bar{P}}}

\def\brtheta{\bar{\theta}}

\newcommand\SDPi{\widehat{\Pi}}

\newcommand\p\partial

\begin{document}

\begin{titlepage}
\title{\vskip -100pt
\vskip 2cm 
Green-Schwarz superstring  on doubled-yet-gauged spacetime}

\author{\sc{Jeong-Hyuck Park}}
\date{}
\maketitle 
\begin{center}
Department of Physics, Sogang University, 35 Baekbeom-ro, Mapo-gu,  Seoul  04107, Korea\\
B. W. Lee Center for Fields, Gravity and Strings, Institute for Basic Science, Daejeon 34047, Korea\\
~\\
\texttt{  park@sogang.ac.kr  }
~\\
~~~\\
\end{center}
\begin{abstract}
\noindent    
We construct a world-sheet action for  Green-Schwarz superstring   in terms of doubled-yet-gauged spacetime coordinates.   For an arbitrarily curved NS-NS background, the action possesses $\mathbf{O}(10,10)$ T-duality, $\mathbf{Spin}(1,9)\times\mathbf{Spin}(9,1)$ Lorentz symmetry, coordinate gauge symmetry,  spacetime doubled-yet-gauged diffeomorphisms,  world-sheet diffeomorphisms and Weyl symmetry. Further, restricted to flat backgrounds,  it  enjoys     maximal  spacetime supersymmetry  and   kappa-symmetry.  After the   auxiliary  coordinate gauge symmetry potential being integrated out,  our action can consistently reduce to the   original undoubled Green-Schwarz action. Thanks to the twofold spin groups, the action is unique: it is specific choices of the NS-NS backgrounds that distinguish   IIA or IIB,  as well as lead to non-Riemannian or non-relativistic superstring \textit{a la~}Gomis-Ooguri which might deserve the  nomenclature,  type IIC.

\end{abstract} 
\thispagestyle{empty}
\end{titlepage}
\newpage
\tableofcontents 




\section{Introduction: \textit{doubled-yet-gauged}}
Ever since General Relativity was established,  it has been customary  to  adopt  Riemannian geometry as  the mathematical  framework  to construct a   theory  of fundamental  physics, such as gravity and string theory.  Accordingly,   the Riemannian metric, $g_{\mu\nu}$, has been privileged to be the only  geometric object which should characterize the nature of  gravity. All other fields are  viewed  as additional `matters' which live on the geometric   background and, at the same time,   source the gravitational field. 

  However,   in string theory,   the metric  is only one   segment  of the massless NS-NS sector which further includes  a two-form gauge potential, $B_{\mu\nu}$, and a scalar dilaton, $\phi$. Under T-duality  the three NS-NS fields transform to each other~\cite{Buscher:1987sk,Buscher:1987qj}. This may well imply an alternative gravitational theory where the whole massless NS-NS sector  becomes geometric  as the gravitational unity. Namely,  the  three  fields, $\{g_{\mu\nu},B_{\mu\nu},\phi\}$, ought to be the  trinity of `stringy  gravity'.  After  series of pioneering works on doubled sigma models~\cite{Duff:1989tf,Tseytlin:1990nb,Tseytlin:1990va,Hull:2004in,Hull:2006qs,Hull:2006va} and Double Field Theory (DFT)~\cite{Siegel:1993xq,Siegel:1993th,Hull:2009mi}, such an idea has been    materialized recently. \\

\noindent First of all,   the number of  the spacetime coordinates is doubled from $D$ to $D{+D}$~\cite{Duff:1989tf}, by adding  dual coordinates, $\tx_{\mu}$, to the  conventional ones,  $x^{\mu}$,  to  form doubled  $(D{+D})$-dimensional coordinates,  
\be
\ba{ll}
x^{M}=(\tx_{\mu},x^{\nu})\,,\quad&\quad M=1,2,3,\cdots,D{+D}\,.
\ea
\ee 
On the doubled coordinate  space, T-duality becomes a run-of-the-mill $\ODD$ rotation.\footnote{Yet, we stress that the doubled coordinates are not restricted to the description of strings but equally  applicable to point-like particle dynamics, see \textit{e.g.~}\cite{Ko:2016dxa}.}  However, despite of the doubling,  the physical  dimension  of the spacetime should be  undoubled: the doubled coordinates must describe $D$-dimensional physics.
One governing  geometric principle,   proposed in \cite{Park:2013mpa} and   pursued in this work,  is   the notion of \textit{doubled-yet-gauged} coordinate system: the doubled coordinate space is gauged  by an equivalence relation, called \textit{coordinate gauge symmetry},
\be
x^{M}\quad\sim\quad x^{M} +\cJ^{MN}\Phi_{\rm s}\partial_{N}\Phi_{\rm t}\,,
\label{CGS}
\ee
such that it is a gauge orbit  that represents a single physical point. Hereafter, $\Phi_{\rm s}, \Phi_{\rm t}$ and $ \Phi_{\rm u}$ denote arbitrary fields and   their arbitrary derivative descendants which must belong to  the theory  employing  the doubled-yet-gauged coordinate system. Further,  $\cJ^{MN}$ is the  inverse of the $\ODD$ invariant metric,
\be
\ba{ll}
\cJ_{MN}=\left(\ba{cc}0&1\\1&0\ea\right)\,, \quad&\quad \cJ^{LM}\cJ_{MN}=\delta^{L}_{~N}\,,
\ea
\label{defcJ}
\ee
which can freely raise and lower the $\ODD$ vector indices, \textit{e.g.~}$\cJ^{MN}\partial_{N}=\partial^{M}$. \\

\noindent In Double Field Theory, the equivalence relation, (\ref{CGS}), is realized   by requiring that all the fields in the theory are invariant under the coordinate gauge symmetry shift,
\be
\ba{ll}
\Phi_{\rm u}(x)=\Phi_{\rm u}(x+\Delta)\,,\quad&\quad\Delta^{M}=\Phi_{\rm s}\partial^{M}\Phi_{\rm t}\,.
\ea
\label{CGSDFT}
\ee
This invariance is then equivalent, \textit{i.e.~}necessary~\cite{Park:2013mpa} and sufficient~\cite{Lee:2013hma}, to   the `{section condition}'~\cite{Siegel:1993th},\footnote{The equivalence basically follows from the power series expansion of ~\eqref{CGSDFT}. It is worth while to note that the former (strong) constraint in \eqref{sectioncon} implies the latter (weak) one, since   $\,\partial_{A}\partial^{B}\Phi_{\rm s}\,\partial_{B}\partial^{C}\Phi_{\rm s}=0$ means that  $\partial_{A}\partial^{B}\Phi_{\rm s}$ is a nilpotent matrix and hence is  traceless.  On the other hand, replacing  $\Phi_{\rm u}$ by the product, $\Phi_{\rm s}\Phi_{\rm t}$,  the latter gives  the former.}
\be
\ba{ll}
\partial_{A}\Phi_{\rm s}\partial^{A}\Phi_{\rm t}=0\,,\quad&\quad
\partial_{A}\partial^{A}\Phi_{\rm u}=0\,,
\ea
\label{sectioncon}
\ee
which are   the differential constraints required  for the consistency of  DFT.\footnote{Yet, \textit{c.f.}~\cite{Geissbuhler:2011mx,Grana:2012rr,Cho:2015lha} for the discussion on alternative constraints. }  Upon the section condition, the generalized Lie derivatives given by~\cite{Siegel:1993th,Hull:2009zb} (\textit{c.f.~}\cite{Gualtieri:2003dx,Grana:2008yw}), 
\be
\hcL_{\cV}T_{M_{1}\cdots M_{n}}:=\cV^{N}\partial_{N}T_{M_{1}\cdots M_{n}}+\omega\partial_{N}\cV^{N}T_{M_{1}\cdots M_{n}}+\sum_{i=1}^{n}(\partial_{M_{i}}\cV_{N}-\partial_{N}\cV_{M_{i}})T_{M_{1}\cdots M_{i-1}}{}^{N}{}_{M_{i+1}\cdots  M_{n}}\,,
\label{tcL}
\ee
are  closed under commutations: 
\be
\ba{cc}
\left[\hcL_{\cU},\hcL_{\cV}\right]=\hcL_{[\cU,\cV]_{\rm{C}}}\,,~~&~~~
{}[\cU,\cV]^{M}_{\rm{C}}:= \cU^{N}\partial_{N}\cV^{M}-\cV^{N}\partial_{N}\cU^{M}+\half \cV^{N}\partial^{M}\cU_{N}-\half\cU^{N}\partial^{M}\cV_{N}\,.
\ea
\ee 
That is to say, the generalized Lie derivative generates the diffeomorphisms on the doubled-yet-gauged  coordinate system (see \cite{Hohm:2012gk,Park:2013mpa,Berman:2014jba,Hull:2014mxa,Chaemjumrus:2015vap,Naseer:2015tia,Rey:2015mba} for finite transformations). Then, in a   parallel manner to Riemannian geometry,  by taking the whole massless NS-NS sector as the geometric fields, the relevant torsion-free  diffeomorphism connection (\textit{i.e.}~``Christoffel symbols"),  covariant derivatives,  a two-indexed curvature (\textit{i.e.~}``Ricci curvature") and a scalar  curvature have been constructed~\cite{Jeon:2011cn} (\textit{c.f.~}\cite{Jeon:2010rw}).\footnote{Yet, there appears no four-indexed (``Riemann")  curvature~\cite{Hohm:2011si}.}   By now, the formalism has been well developed, such that ${D=10}$ maximally supersymmetric DFT has been constructed to the full order in fermions~\cite{Jeon:2012hp},  and the   Standard Model itself has been `double-field-theorized' to covariantly couple   to the massless NS-NS sector of the gravitational DFT~\cite{Choi:2015bga}  (\textit{c.f.}~\cite{Jeon:2011kp,Jeon:2011vx,Jeon:2011sq,Hohm:2011zr,Jeon:2012kd} for related earlier works). In particular, the maximally supersymmetric DFT  not only contains and unifies type IIA and IIB supergravities but can also  feature     `non-Riemannian'   geometry, as we review below. 

The massless NS-NS sector enters  (bosonic) DFT  in the form of a symmetric $\ODD$ element,  called ``generalized metric",
\be
\ba{ll}
\cH_{MN}=\cH_{NM}\,,\quad&\quad \cH_{K}{}^{L}\cH_{M}{}^{N}\cJ_{LN}=\cJ_{KM}\,,
\ea
\label{defDFTmetric}
\ee
along with a scalar density, $e^{-2d}$, having the weight of unity. Combined with the $\ODD$ invariant metric, the generalized metric can produce a pair of orthogonal and complete  symmetric projectors,
\be
\ba{lll}
P_{MN}=P_{NM}=\half(\cJ_{MN}+\cH_{MN})\,,\quad&\quad
P_{L}{}^{M}P_{M}{}^{N}=P_{L}{}^{N}\,,\quad&\quad
P_{K}{}^{L}\brP_{L}{}^{M}=0\,,\\
\brP_{MN}=\brP_{NM}=\half(\cJ_{MN}-\cH_{MN})\,,\quad&\quad
\brP_{L}{}^{M}\brP_{M}{}^{N}=\brP_{L}{}^{N}\,,\quad&\quad
P_{MN}+\brP_{MN}=\cJ_{MN}\,.
\ea
\label{projectors}
\ee
These $\ODD$ covariant variables may be generically parametrized    in terms of  the conventional variables, $\{g_{\mu\nu},B_{\mu\nu},\phi\}$, but there are also exceptions which do not allow such parametrization  even locally at all. This leads  to the notion of  `non-Riemannian' backgrounds~\cite{Lee:2013hma,Ko:2015rha} (\textit{c.f.~}\cite{Gomis:2000bd}).

The unification of IIA  and IIB is  due to the facts that \textit{i)} the local Lorentz  spin group in DFT is twofold, $\SpinD\times\oSpinD$ (basically one for $P_{MN}$ and the other for $\brP_{MN}$),        \textit{ii)} the maximally supersymmetric DFT is chiral with respect to both spin groups, $\Spint$ and $\oSpint$,  \textit{iii)} hence,  the theory is  unique: it admits IIA, IIB and non-Riemannian backgrounds as different types of solutions. In this sense,   the last type might deserve the nomenclature, type IIC.  \\

\noindent On the other hand, in doubled sigma models where the doubled coordinates are dynamical,  the coordinate gauge symmetry~(\ref{CGS})  calls for     the relevant   gauge  connection rather explicitly~\cite{Lee:2013hma},
\be
DX^{M}:=\rd X^{M}-\cA^{M}\,.
\ee 
As in any gauge theory, the  gauge  potential, $\cA^{M}$, should meet precisely the same property as the gauge generator which is,  in the present case, $\Delta^{M}$ in \eqref{CGSDFT}. Hence, similarly  to the section condition~\eqref{sectioncon},  the coordinate gauge symmetry potential satisfies 
\be
\ba{ll}
\cA^{M}\partial_{M}=0\,,\quad&\quad
\cA^{M}\cA_{M}=0\,.
\ea
\label{secconA}
\ee
Respecting  these constraints, 
the coordinate gauge symmetry is  realized as
\be
\ba{lll}
\deltaCG X^{M}=\Phi_{\rm s}\partial^{M}\Phi_{\rm t}\,,\quad&\quad
\deltaCG\cA^{M}=\rd\left(\Phi_{\rm s}\partial^{M}\Phi_{\rm t}\right)\,,\quad&\quad\deltaCG (DX^{M})=0\,.
\ea
\label{deltaCG}
\ee
Further, while $\rd X^{M}$ is not a diffeomorphism covariant vector, $DX^{M}$ is so:
\be
\ba{lll}
\deltaV X^{M}=\cV^{M}\,,\quad&\quad
\deltaV (\rd X^{M})=\rd X^{N}\partial_{N}\cV^{M}\,,\quad&\quad
\deltaV (DX^{M})=(\partial_{N}\cV^{M}-\partial^{M}\cV_{N})D X^{N}\,,\\
\multicolumn{3}{c}{
\deltaV\cA^{M}=
(\partial_{N}\cV^{M}-\partial^{M}\cV_{N})\cA^{N}+\partial^{M}\cV_{N}\rd X^{N}=-\partial^{M}\cV_{N}\cA^{N}+\partial^{M}\cV_{N}\rd X^{N}\,.}
\ea
\label{deltaU}
\ee
It is this  gauged one-form, $DX^{M}$, with the obvious kinetic term, $DX^{M}DX^{N}\cH_{MN}(X)$,  that can be used to construct $\ODD$ T-duality, diffeomorphisms and coordinate gauge symmetry  covariant sigma models: 
\begin{itemize}
\item[\textit{i)}] world-sheet action for a string~\cite{Lee:2013hma}, 
\be
\cS_{\scriptscriptstyle{\rm{string}}}={\textstyle{\frac{1}{4\pi\alpha^{\prime}}}}{\dis{\int}}\rd^{2}\sigma~\Big[\,-\half\sqrt{-h}h^{ij}\rdA_{i}X^{M}\rdA_{j}X^{N}\cH_{MN}(X)-\epsilon^{ij}\rdA_{i}X^{M}\cA_{jM}\,\Big]\,,
\label{bosonicaction}
\ee 
\item[\textit{ii)}] world-line action for a point-like particle~\cite{Ko:2016dxa}, 
\be\dis{
\cS_{\scriptscriptstyle{\rm{particle}}}=
\int\rd\tau ~\Big[\,e^{-1\,}D_{\tau}X^{M}D_{\tau}X^{N}\cH_{MN}(X)-\quarter m^{2}e\,\Big]\,.}
\label{particleaction}
\ee 
\end{itemize}
The former result~\eqref{bosonicaction} was essentially a re-derivation of  the doubled string action proposed by   Hull~\cite{Hull:2006va}, with the   coordinate gauge symmetry interpretation added.  Especially upon Riemannian backgrounds, the Euler-Lagrangian  equation of  the   coordinate gauge symmetry potential,  $\cA_{iM}$,  implies the self-duality (\textit{i.e.~}chirality)  over the entire doubled spacetime, \textit{c.f.~}\eqref{SD2},
\be
\rdA_{i}X_{M}+\textstyle{\frac{1}{\sqrt{-h}}}\epsilon_{i}{}^{j}\cH_{M}{}^{N}\rdA_{j}X_{N}=0\,,
\ee
and the Euler-Lagrangian  equation of $X^{M}$ gets simplified to give   the \textit{stringy geodesic equation},
\be
\textstyle{\frac{1}{\sqrt{-h}}}\partial_{i}(\sqrt{-h}\rdA^{i}X^{M}\cH_{ML})+\Gamma_{LMN}(\brP\rdA_{i}X)^{M}(P\rdA^{i}X)^{N}=0\,,
\ee
where $\Gamma_{LMN}$ is the stringy Christoffel connection  obtained in \cite{Jeon:2011cn}, and $(\brP\rdA_{i}X)^{M}=\brP^{M}{}_{N}\rdA_{i}X^{N}$ \textit{etc}. It is worth while to note that  the world-sheet topological term in \eqref{bosonicaction}  transforms to total derivatives      under the coordinate gauge symmetry~\eqref{deltaCG} as well as  under the  diffeomorphisms~\eqref{deltaU}~\cite{Lee:2013hma}, 
\be
\ba{ll}
\deltaCG\!\left(\epsilon^{ij}\rdA_{i}X^{M}\cA_{jM}\right)
=-\partial_{i}\left(\epsilon^{ij}\Phi_{\rm s}\partial_{j}\Phi_{\rm t}\right)\,,
\quad&\quad\deltaV\!\left(\epsilon^{ij}\rdA_{i}X^{M}\cA_{jM}\right)=\partial_{i}
\left(\epsilon^{ij}\cV_{M}\partial_{j}X^{M}\right)\,.
\ea
\ee
The kinetic terms in \eqref{bosonicaction} and \eqref{particleaction} are invariant under the coordinate gauge symmetry, and they transform `covariantly' under  the  diffeomorphisms,  $\deltaV\left(DX^{M}DX^{N}\cH_{MN}\right)
=DX^{M}DX^{N}\hcL_{\cV}\cH_{MN}$, such that any Killing vector satisfying ${\hcL_{\cV}\cH_{MN}=0}$  induces  a  Noether symmetry of the action.

In the above doubled sigma models, the  gauge potentials are all  auxiliary. After they are integrated out, the doubled sigma models consistently reduce to the conventional  undoubled   string and  particle  actions. \\

\noindent It is the purpose of the present paper  to  supersymmetrize the above doubled string action~\eqref{bosonicaction}, or equivalently  to formulate the renowned  Green-Schwarz superstring action~\cite{Green:1983wt} on the doubled-yet-gauged spacetime,   as the complementary world-sheet counterpart  to the  maximally supersymmetric  DFT~\cite{Jeon:2012hp}.

Our constructed action is going to be  symmetric, with respect to
\begin{itemize}
\item $\Ott$ T-duality,
\item  $\Spint\times\oSpint$ Lorentz symmetry,
\item coordinate gauge symmetry,
\item target-spacetime doubled-yet-gauged  diffeomorphisms (over Killing directions),
\item world-sheet diffeomorphisms,
\item conformal symmetry under Weyl transformations,
\end{itemize}
and, in addition, restricted to flat NS-NS backgrounds,
\begin{itemize}
\item target-spacetime $16{+16}$ global supersymmetry,
\item $16{+16}$ kappa-symmetry.
\end{itemize}
 Since we do not include spin connections, the $\Spint\times\oSpint$ Lorentz symmetry is going to be global rather than local.  Nevertheless, the global twofold spin structure ensures to unify IIA and IIB superstrings: different choices of the NS-NS backgrounds give rise to  IIA or IIB, as well as  non-Riemannian IIC superstrings.   Once again, after  the auxiliary coordinate gauge symmetry potential being integrated out, our action  reduces  consistently to the Green-Schwarz type IIA/B superstring action if the background is Riemannian. Alternatively, upon a non-Riemannian background, our action leads to the supersymmetric  extension of the Gomis-Ooguri non-relativistic string~\cite{Gomis:2000bd}. \\

\noindent For further inspiring   precursors,  we refer readers  to \cite{Hull:2006va,Driezen:2016tnz} for the world-sheet supersymmetries, \cite{Hatsuda:2015cia} for the  construction of chiral affine (super-)Lie algebras,   \cite{Cederwall:2016ukd}
 for  the   T-duality  supergroup, $\mbox{OSp}(D,D|2s)$,  as well as  \cite{Blair:2013noa} for a doubled  Hamiltonian sigma model and \cite{Freidel:2013zga,Freidel:2015pka} for the Born reciprocity. We also refer the work by Bandos~\cite{Bandos:2015cha,Bandos:2016jez} on the construction of a PST superstring action in doubled superspace. ~\\



\section{Green-Schwarz superstring in terms of  doubled-yet-gauged coordinates}
In this section, firstly we present our main result, \textit{i.e.~}`the construction of the Green-Schwarz superstring action on the doubled-yet-gauged spacetime',  and then provide the relevant explanations, such as the conventions, the field contents, the target-spacetime supersymmetry and the kappa-symmetry. The reductions to the undoubled type IIA, IIB and non-relativistic IIC superstrings will be discussed in the next section.\\

\subsection{Main result}

We propose the Green-Schwarz superstring action on the doubled-yet-gauged spacetime, 
\be
\cS_{\scriptscriptstyle{\rm{superstring}}}={\textstyle{\frac{1}{4\pi\alpha^{\prime}}}}{\dis{\int}}\rd^{2}\sigma~\cL_{\scriptscriptstyle{\rm{superstring}}}\,,
\label{Action}
\ee
with the Lagrangian,
\be
\cL_{\scriptscriptstyle{\rm{superstring}}}=-\half\sqrt{-h}h^{ij}\Pi_{i}^{M}\Pi_{j}^{N}\cH_{MN}
-\epsilon^{ij}D_{i}X^{M}\left(\cA_{jM}-i\Sigma_{jM}\right)\,.
\label{Lagrangian}
\ee
Here, equipped with the map from the  string world-sheet to the   doubled-yet-gauged  target-spacetime, 
\be
\ba{lll}
\sigma^{i}&\longrightarrow&
X^{M}=\left(\widetilde{X}_{\mu}\,,X^{\nu}\right)\,,
\ea
\ee
and a pair of Majorana-Weyl spinors,  $\theta^{\alpha}$ for $\Spint$ and $\thetap{}^{\bralpha}$ for $\oSpint$,  
we set
\be
\ba{ll}
\Pi_{i}^{M}:=\rdA_{i}X^{M}-i\Sigma_{i}^{M}\,,\quad&\quad
\Sigma_{i}^{M}:=
\brtheta\gamma^{M}\partial_{i}\theta+\brthetap\brgamma^{M}\partial_{i}\thetap\,.
\ea
\ee

For an arbitrarily curved NS-NS background, the action possesses  the manifest $\Ott$ T-duality,  the ${\Spint\times\oSpint}$ global Lorentz symmetry, the coordinate gauge symmetry, the target-spacetime doubled-yet-gauged diffeomorphisms over any Killing direction, world-sheet diffeomorphisms and  Weyl symmetry.        

  Moreover, when the background is flat, the action is invariant under $16{+16}$ global target-spacetime supersymmetry,
\be
\ba{lllll}
\deltaS X^{M}=i\brvarepsilon\gamma^{M}\theta+i\brvarepsilonp\brgamma^{M}\thetap\,,\quad&\quad
\deltaS\theta=\varepsilon\,,\quad&\quad
\deltaS\thetap=\varepsilonp\,,\quad&\quad
\deltaS h_{ij}=0\,,\quad&\quad
\deltaS\cA_{iM}=0\,,
\ea
\label{SUSY}
\ee
as well as  $16{+16}$ local fermionic kappa-symmetry,
\be
\ba{lll}
\deltak X^{M}=i\brtheta\gamma^{M}\deltak\theta+i\brthetap\brgamma^{M}\deltak\thetap\,,\quad&\quad
\deltak\theta=h_{+}^{ij}\Pi_{iM}\gamma^{M}\kappa_{j}\,,\quad&\quad
\deltak\thetap=h_{-}^{ij}\Pi_{iM}\brgamma^{M}\kappap_{j}\,,\\
\multicolumn{3}{c}{
\deltak(\sqrt{-h}h^{ij})=-8i\sqrt{-h}(h_{+}^{ik}h_{+}^{jl}\partial_{k}\brtheta\kappa_{l}+h_{-}^{ik}h_{-}^{jl}
\partial_{k}\brthetap\kappap_{l})\,,}\\
\multicolumn{3}{c}{\deltak\cA_{iM}=-2i\left(h_{+i}{}^{j}h_{+}^{kl}\partial_{j}\brtheta\kappa_{l}
+h_{-i}{}^{j}h_{-}^{kl}\partial_{j}\brthetap\kappap_{l}
\right){\bf{\Big[}}\SDPi_{kM}{\bf{\Big]}}_{\scriptscriptstyle\rm{\bf projected}}\,.}
\ea
\label{kappa}
\ee
In the above,  we set a pair of  world-sheet projection matrices,
\be
\ba{ll}
h_{+}^{ij}:=\half\left(h^{ij}+\textstyle{\frac{\epsilon^{ij}}{\sqrt{-h}}}\right)\,,\quad&\quad
h_{-}^{ij}:=\half\left(h^{ij}-\textstyle{\frac{\epsilon^{ij}}{\sqrt{-h}}}\right)=h_{+}^{ji}\,,
\ea
\label{defhpm}
\ee
and the self-dual part of $\Pi_{i}^{M}$,
\be
\SDPi_{i}^{M}:=\Pi_{i}^{M}+\textstyle{\frac{\epsilon_{i}{}^{j}}{\sqrt{-h}}}\cH^{M}{}_{N}\Pi_{j}^{N}\,.
\label{SD}
\ee
Further, ${\bf{\Big[}}\SDPi_{kM}{\bf{\Big]}}_{\scriptscriptstyle\rm{\bf projected}}$  means the projection of $\SDPi_{kM}$ to the coordinate gauge symmetry value, such that
\be
{\bf{\Big[}}\SDPi_{k}^{M}{\bf{\Big]}}_{\scriptscriptstyle\rm{\bf projected}}\times\partial_{M}=0\,.
\ee
Concretely,   without loss of generality up to $\Ott$ rotations,  if we choose the section as
\be
\ba{ll}
\partial_{M}=\left(\tpartial^{\mu}\,,\partial_{\nu}\right)\equiv\left(0\,,\partial_{\nu}\right)\,,\quad&\quad\cA_{iM}\equiv\big(0\,,A_{i\mu}\big)\,,
\ea
\label{sectionchosen}
\ee
we have
\be
{\bf{\Big[}}\SDPi_{i}^{M}{\bf{\Big]}}_{\scriptscriptstyle\rm{\bf projected}}=\left[\left(\widetilde{\SDPi}_{i\mu}\,,\SDPi_{i}^{\nu}\right)\right]_{\scriptscriptstyle\rm{\bf projected}}\equiv
\left(\widetilde{\SDPi}_{i\mu}\,,\,0\right)\,,
\ee
and thus,  
\be
\deltak \cA_{iM}=\big(0\,,\deltak A_{i\mu}\big)=-2i\left(h_{+i}{}^{j}h_{+}^{kl}\partial_{j}\brtheta\kappa_{l}
+h_{-i}{}^{j}h_{-}^{kl}\partial_{j}\brthetap\kappap_{l}
\right)\times\left(0\,,\widetilde{\SDPi}_{k\mu}\right)\,.
\ee
Surely, $\varepsilon$, $\varepsilonp$ are constant Majorana-Weyl  $\Spint$, $\oSpint$ spinors having the same chiralities as $\theta$, $\thetap$ respectively, while $\kappa_{i}$, $\kappap_{j}$ are local Majorana-Weyl spinors with the opposite chiralities.  As stressed by Hull~\cite{Hull:2006va},  the string tension on a doubled space should be halved, \textit{i.e.~}$(4\pi\alphap)^{-1}$. Further explanations are in order  in the following  subsections. \\

\subsection{Conventions and field contents}

Our conventions,  especially for the indices, are identical to \cite{Jeon:2012kd,Jeon:2012hp} and summarized in Table~\ref{TABindices}.

{\scriptsize{\begin{table}[h]
\begin{center}
\begin{tabular}{|c|c|c|}
\hline
Index~&\mbox{~}Representation\mbox{~}&Raising \& Lowering Indices\\
\hline
$M,N,\cdots$&Target-space  diffeomorphism \& $\Ott$ vector&$\cJ_{AB}$\quad\eqref{defcJ}\\
$p,q,\cdots$&$\Spint$  vector~&$\eta_{pq}=\mbox{diag}(-++\cdots+)$ \\
$\alpha,\beta,\cdots$&$\Spint$  spinor~&$C_{\alpha\beta}=C_{\beta\alpha}$\,,\mbox{~~}~$(\gamma^{p})^{T}=C\gamma^{p}C^{-1}$\quad\eqref{chargecon}\\
$\brp,\brq,\cdots$&$\oSpint$  vector~&$\breta_{\brp\brq}=\mbox{diag}(+--\cdots-)$ \\
$\bralpha,\brbeta,\cdots$&$\oSpint$  spinor~&$\brC_{\bralpha\brbeta}=\brC_{\brbeta\bralpha}$\,,    \mbox{~~}~$(\brgamma^{\brp})^{T}=\brC\brgamma^{\brp}\brC^{-1}$\quad\eqref{chargecon}\\
$i,j,\cdots$&World-sheet diffeomorphism vector & $h_{ij}$\\
\hline
\end{tabular}
\caption{Convention of the indices and the corresponding ``metric" to raise or lower the positions.  } 
\label{TABindices}
\end{center}
\end{table}}}

Since the Spin group is twofold as $\Spint\times\oSpint$, there exist a pair of gamma matrices,  $(\gamma^{p})^{\alpha}{}_{\beta}$ and  $(\brgamma^{\brp})^{\bralpha}{}_{\brbeta}$. The corresponding  charge conjugation matrices, $C_{\alpha\beta}$ and $\brC_{\bralpha\brbeta}$, satisfy for $n=0,1,2,\cdots$,
\be
\ba{ll}
\left(C\gamma^{p_{1}p_{2}\cdots p_{n}}\right)_{\alpha\beta}=(-1)^{n(n-1)/2}\left(C\gamma^{p_{1}p_{2}\cdots p_{n}}\right)_{\beta\alpha}\,,~~&~~
\left(\brC\brgamma^{\brp_{1}\brp_{2}\cdots \brp_{n}}\right)_{\bralpha\brbeta}=(-1)^{n(n-1)/2}\left(\brC\brgamma^{\brp_{1}\brp_{2}\cdots \brp_{n}}\right)_{\brbeta\bralpha}\,.
\ea
\label{chargecon}
\ee
A  well-known  crucial Fierz identity is
\be
\ba{ll}
\left(C\gamma^{p}\gamma_{+}\right)_{(\alpha\beta}
\left(C\gamma_{p}\gamma_{+}\right)_{\gamma)\delta}=0\,,
\quad&\quad
\left(\brC\brgamma^{\brp}\brgamma_{+}\right)_{(\bralpha\brbeta}
\left(\brC\brgamma_{\brp}\brgamma_{+}\right)_{\brgamma)\brdelta}=0\,,
\ea
\label{Fierz}
\ee
where  $\gamma_{+}$, $\brgamma_{+}$ denote the usual chiral projection matrices,
\be
\ba{llll}
\gamma_{+}:=\half\left[1+\gamma^{\eleven}\right]\,,\quad&\quad\gamma^{\eleven}:=\gamma^{012\cdots 9}\,,\quad&\quad(\gamma_{+})^{2}=\gamma_{+}\,,\quad&\quad(\gamma^{\eleven})^{2}=1\,,\\
\brgamma_{+}:=\half\left[1+\brgamma^{\eleven}\right]\,,\quad&\quad\brgamma^{\eleven}:=\brgamma^{012\cdots 9}\,,\quad&\quad(\brgamma_{+})^{2}=\brgamma_{+}\,,\quad&\quad(\brgamma^{\eleven})^{2}=1\,.
\ea
\label{gammae}
\ee

The NS-NS background of the action is given by the DFT-vielbeins satisfying  four   \textit{defining properties}:
\be
\ba{llll}
V_{Mp}V^{M}{}_{q}=\eta_{pq}\,,\quad&\quad
\brV_{M\brp}\brV^{M}{}_{\brq}=\breta_{\brp\brq}\,,\quad&\quad
V_{Mp}\brV^{M}{}_{\brq}=0\,,\quad&\quad V_{Mp}V_{N}{}^{p}+\brV_{M\brp}\brV_{N}{}^{\brp}=\cJ_{MN}\,.
\ea
\label{defV}
\ee
That is to say, they are normalized, orthogonal and complete.  They correspond to the ``square-roots" of the projectors~\eqref{projectors}, as 
\be
\ba{ll}
P_{MN}=V_{Mp}V_{N}{}^{p}\,,\quad&\quad\brP_{MN}=\brV_{M\brp}\brV_{N}{}^{\brp}\,,
\ea
\ee 
while the generalized metric is given by the difference, \be
\cH_{MN}=V_{Mp}V_{N}{}^{p}-\brV_{M\brp}\brV_{N}{}^{\brp}\,.
\ee  
It follows then that 
\be
\ba{ll}
\cH_{M}{}^{N}V_{Np}=+V_{Mp}\,,\quad&\quad
\cH_{M}{}^{N}\brV_{N\brp}=-\brV_{M\brp}\,.
\ea
\label{DFTmetricVV}
\ee  
In this way, the DFT-vielbeins  simultaneously diagonalize $\cJ_{MN}$ and $\cH_{MN}$ into `$\mbox{diag}(\eta,\breta)$' and `$\mbox{diag}(\eta,-\breta)$' respectively.  
As a solution to  (\ref{defV}),  they may be   parametrized    generically by ordinary    zehnbeins  and $B$-field~\eqref{Vform1},   (up to  field redefinitions, \textit{e.g.~}\cite{Andriot:2013xca}); or they may not admit such a   conventional \textit{i.e.~}Riemannian parametrization~\cite{Lee:2013hma,Ko:2015rha}.

Contracted with the DFT-vilebeins, the gamma matrices  can carry  $\Ott$ vector indices,   such as $\gamma^{M}=V^{M}{}_{p}\gamma^{p}$ and $\brgamma^{M}=\brV^{M}{}_{\brp}\brgamma^{\brp}$ which satisfy then
\be
\ba{ll}
\gamma^{M}\gamma^{N}+\gamma^{N}\gamma^{M}=2P^{AB}\,,\quad&\quad
\brgamma^{M}\brgamma^{N}+\brgamma^{N}\brgamma^{M}=2\brP^{AB}\,.
\ea
\label{CliffordPP}
\ee

All the spinors  are fermionic    Majorana-Weyl spinors for either the  $\Spint$ or the $\oSpint$ Lorentz group,  in particular to meet
\be
\ba{llll}
\theta=\gamma^{\eleven}\theta\,,\quad&\quad
\brtheta=\theta^{T}C=-\brtheta\gamma^{\eleven}\,,\quad&\quad\thetap=\brgamma^{\eleven}\thetap\,,\quad&\quad
\brthetap=\thetap{}^{T}\brC=-\brthetap\brgamma^{\eleven}\,.
\ea
\ee

It is worth while to note that, using the properties of the coordinate gauge symmetry potential~\eqref{secconA}, we may rewrite  the world-sheet topological term as
\be
\epsilon^{ij}D_{i}X^{M}\left(\cA_{jM}-i\Sigma_{jM}\right)=
\epsilon^{ij}\left(\Pi_{i}^{M}\cA_{jM}-i\partial_{i}X^{M}\Sigma_{jM}\right)\,.
\ee


\subsection{Target-spacetime supersymmetry and Wess-Zumino term}
For flat NS-NS backgrounds where the DFT-vielbeins are all constant, $\Pi_{i}^{M}$ is 
 target-spacetime supersymmetry   invariant,  under \eqref{SUSY},
\be
\deltaS\Pi_{i}^{M}=0\,,
\label{SUSYPi}
\ee
and the Lagrangian transforms to total derivatives,  implying the  invariance of the action,
\be
\ba{ll}
\!\deltaS\cL_{\scriptscriptstyle{\rm{superstring}}}&
=\epsilon^{ij}\!\left(\partial_{i}X^{M}\partial_{j}\deltaS X_{M}+\Sigma_{i}^{M}\deltaS\Sigma_{jM}\right)\\
{}&=-\epsilon^{ij}\partial_{i}\!\left(\partial_{j}X^{M}\deltaS X_{M}
+\textstyle{\frac{1}{3}}\brvarepsilon\gamma^{p}\theta\brtheta\gamma_{p}\partial_{j}\theta
+\textstyle{\frac{1}{3}}\brvarepsilonp\brgamma^{\brp}\thetap\brthetap\brgamma_{\brp}\partial_{j}
\thetap\right)\,.
\ea
\ee
In the above, the second equality  follows essentially from the Fierz identity~(\ref{Fierz}) which enables us  to write
\be
\ba{ll}
\epsilon^{ij}\gamma^{p}\partial_{i}\theta\brtheta\gamma_{p}\partial_{j}\theta
=\textstyle{\frac{1}{3}}\partial_{i}\left(\epsilon^{ij}\gamma^{p}\theta\brtheta\gamma_{p}
\partial_{j}\theta\right)\,,\quad&\quad
\epsilon^{ij}\brgamma^{\brp}\partial_{i}\thetap\brthetap\brgamma_{\brp}
\partial_{j}\thetap=\textstyle{\frac{1}{3}}\partial_{i}\left(\epsilon^{ij}\brgamma^{\brp}\thetap\brthetap
\brgamma_{\brp}\partial_{j}\thetap\right)\,.
\ea
\label{Fierztd}
\ee
~\\

In fact, extending  the two-dimensional world-sheet to a fictitious  three-dimensional space and using identities  due to  \eqref{Fierz} like
\be
\ba{ll}
\epsilon^{ijk}
\brtheta\gamma^{p}\partial_{i}\theta\,\partial_{j}\brtheta\gamma_{p}\partial_{k}\theta=0\,,\quad&\quad
\epsilon^{ijk}
\brthetap\brgamma^{\brp}\partial_{i}\thetap\,\partial_{j}\brthetap\brgamma_{\brp}\partial_{k}\thetap=0\,,
\ea
\ee
we may straightforwardly compute the `exterior derivative' of the topological term,
\be
\ba{ll}
\epsilon^{ijk}\partial_{k}\big[D_{i}X^{M}\left(\cA_{jM}-i\Sigma_{jM}\right)\big]=\epsilon^{ijk}
\Pi_{i}^{M}\big(
i\partial_{j}\brtheta\gamma_{M}\partial_{k}\theta+i
\partial_{j}\brthetap\brgamma_{M}\partial_{k}\thetap
-\half\cF_{jk M}\big)\,,
\ea
\label{WZ}
\ee
where we  set the field strength of the coordinate gauge symmetry potential, $\cF_{jk M}:=\partial_{j}\cA_{kM}-\partial_{k}\cA_{jM}$.  
The resulting `three-form' on the right hand side of the equality in \eqref{WZ}  then corresponds to the  Wess-Zumino term~\cite{Wess:1971yu}  for Green-Schwarz superstring~\cite{Henneaux:1984mh} now   on   doubled-yet-gauged spacetime. As desired, it  is manifestly invariant under the global target-spacetime supersymmetry~\eqref{SUSY}.


\subsection{Fermionic kappa-symmetry}

For the systematic derivation  of the kappa-symmetry, we start with generic variations of the spinors, $\delta\theta$, $\delta\thetap$,  and the auxiliary fields,  $\delta \cA_{iM}$, $\delta h_{ij}$, while we set, with the opposite sign compared to the target-spacetime supersymmetry~\eqref{SUSY},
\be
\delta X^{M}=i\brtheta\gamma^{M}\delta\theta+i\brthetap\brgamma^{M}\delta\thetap
=-i\delta\brtheta\gamma^{M}\theta-i\delta\brthetap\brgamma^{M}\thetap\,.
\label{deltaX}
\ee 
It follows straightforwardly upon flat backgrounds, 
\be
\delta\Pi_{i}^{M}
=-2i\delta\brtheta\gamma^{M}\partial_{i}\theta-2i\delta\brthetap
\brgamma^{M}\partial_{i}\thetap-\delta\cA_{i}^{M}\,,
\ee 
and the kinetic term transforms as
\be
\ba{l}
\delta\left(-\half\sqrt{-h}h^{ij}\Pi_{i}^{M}\Pi_{j}^{N}\cH_{MN}\right)\\
=-\half\delta\left(\sqrt{-h}h^{ij}\right)\Pi_{i}^{M}\Pi_{j}^{N}\cH_{MN}+
\sqrt{-h}h^{ij}\Pi_{iM}\left(2i\delta\brtheta\gamma^{M}\partial_{j}\theta-
2i\delta\brthetap\brgamma^{M}\partial_{j}\thetap+\cH^{MN}\delta\cA_{jN}\right)\,.
\ea
\label{0deltaL1}
\ee
On the other hand, the Fierz identity~\eqref{Fierztd} implies for  arbitrary $\delta\theta$ and $\delta\thetap$,
\be
\ba{lll}
\epsilon^{ij}\partial_{i}(\brtheta\gamma^{p}\delta\theta)\brtheta\gamma_{p}\partial_{j}\theta&=&
\partial_{i}\left(\epsilon^{ij}\brtheta\gamma^{p}\delta\theta\brtheta
\gamma_{p}\partial_{j}\theta\right)+
2\epsilon^{ij}\delta\brtheta\gamma^{p}\partial_{i}\theta\brtheta
\gamma_{p}\partial_{j}\theta\,,\\
\epsilon^{ij}\partial_{i}(\brthetap\brgamma^{\brp}\delta\thetap)\brthetap\brgamma_{\brp}\partial_{j}\thetap
&=&
\partial_{i}\left(\epsilon^{ij}\brthetap\brgamma^{\brp}
\delta\thetap\brthetap\brgamma_{\brp}\partial_{j}\thetap\right)
+2\epsilon^{ij}\delta\brthetap\brgamma^{\brp}
\partial_{i}\thetap\brthetap\brgamma_{\brp}\partial_{j}\thetap\,,
\ea
\label{Fierztd2}
\ee
which in turn enable us to organize  the variation of the world-sheet topological term as
\be
\ba{l}
\delta\left[-\epsilon^{ij}D_{i}X^{M}\left(\cA_{jM}-i\Sigma_{jM}\right)\right]\\
=2i\epsilon^{ij}\Pi_{i}{}^{M}\left(\delta\brtheta\gamma_{M}\partial_{j}\theta+
\delta\brthetap\brgamma_{M}\partial_{j}\thetap\right)+\epsilon^{ij}\delta\cA_{iM}\Pi_{j}^{M}
-\partial_{i}\left[\epsilon^{ij}\left(\Pi_{j}{}^{M}+\cA_{j}^{M}\right)\delta X_{M}\right]\,.
\ea
\label{0deltaL2}
\ee
Combining \eqref{0deltaL1} and \eqref{0deltaL2}, with \eqref{defhpm}, \eqref{SD},  we obtain  
\be
\ba{ll}
\delta\cL_{\scriptscriptstyle{\rm{superstring}}}=&-\half\delta\left(\sqrt{-h}h^{ij}\right)\Pi_{i}^{M}\Pi_{j}^{N}\cH_{MN}\,+\,4i\sqrt{-h}\Pi_{iM}\left(h_{+}^{ij}\delta\brtheta\gamma^{M}\partial_{j}\theta-
h_{-}^{ij}\delta\brthetap\brgamma^{M}\partial_{j}\thetap\right) \\
{}&+\epsilon^{ij}\delta\cA_{iM}\SDPi_{j}^{M}\,-\,\partial_{i}\left[\epsilon^{ij}\left(\Pi_{j}{}^{M}+\cA_{j}^{M}\right)\delta X_{M}\right]\,,
\ea
\label{deltaL}
\ee
where the world-sheet projection matrices, $\h_{\pm}^{ij}$~\eqref{defhpm}, naturally appear. They satisfy
\be
\ba{llll}
h_{+j}^{i}h_{+k}^{j}=h^{i}_{+k}\,,\quad&\quad
h_{-j}^{i}h_{-k}^{j}=h^{i}_{-k}\,,\quad&\quad
h_{+j}^{i}h_{-k}^{j}=0\,,\quad&\quad
h_{+j}^{i}+h_{-j}^{i}=\delta^{i}_{~j}\,.
\ea
\label{hpmprop}
\ee
There are four terms on the right hand side of the  equality in \eqref{deltaL}. The last term is total derivative and hence harmless. The first term is quadratic in $\Pi_{iM}$ and needs to be canceled by other two terms (\textit{i.e.~}second and third). For this, the variations of the fermions  need to be linear in $\Pi_{i}^{M}$, such as
\be
\ba{ll}
\deltak\theta=\Pi_{iM}\gamma^{M}\zeta^{i}\,,\quad&\quad\deltak\thetap=\Pi_{iM}\brgamma^{A}\zeta^{\prime i}\,.
\ea
\label{deltatheta}
\ee
Substituting this  ansatz into \eqref{deltaL},  from \eqref{ehprop1}, \eqref{ehprop2},  (\ref{gamma2e}) and through an intermediate step~\eqref{AdeltaL2}, the variation of the Lagrangian further reduces to
\be
\ba{lll}
\delta\cL_{\scriptscriptstyle{\rm{superstring}}}&=&-2i\left(h_{-j}^{i}\partial_{i}\brtheta\gamma_{MN}\zeta^{j}+h_{+j}^{i}\partial_{i}
\brthetap\brgamma_{MN}\zetap{}^{j}\right)
\left(\epsilon^{kl}\Pi_{k}^{M}\Pi_{l}^{N}\right)\\
{}&{}&-\half
\left[\delta\left(\sqrt{-h}h^{ij}\right)+8i\sqrt{-h}\left(\partial_{k}\brtheta\zeta^{(i}h_{+}^{j)k}+
\partial_{k}\brthetap\zetap{}^{(i}h_{-}^{j)k}\right)\right]\Pi_{i}^{M}\Pi_{j}^{N}\cH_{MN}\\
{}&{}&+\epsilon^{ij}\Big[
\delta\cA_{iM}+2i\left(h_{+i}{}^{k}\partial_{k}\brtheta\zeta^{l}+h_{-i}{}^{k}\partial_{k}\brthetap\zetap^{l}\right)\Pi_{lM}
\Big]\SDPi_{j}^{M}\\
{}&{}&-\partial_{i}\left[\epsilon^{ij}\left(\Pi_{j}{}^{M}+\cA_{j}^{M}\right)\delta X_{M}\right]\,.
\ea
\label{deltaL2}
\ee
Except the last harmless term, each line on the right hand side should vanish by itself.  The vanishing of the first line requires
\be
\ba{lllll}
h^{i}_{-j}\zeta^{j}=0\,,~~&~~ h^{i}_{+j}\zetap^{j}=0\quad&~~\Longleftrightarrow~~&\quad
\zeta^{i}=h^{ij}_{+}\kappa_{j}\,,~~&~~ \zetap^{i}=h^{ij}_{-}\kappap_{j}\,,
\ea
\label{kzeta}
\ee
which fix the kappa-symmetry transformations of the fermions, $\deltak\theta,\deltak\thetap$, completely as \eqref{kappa}. Consequently the second line determines the variation of the world-sheet metric~\eqref{kappa}, up to Weyl transformations, which we rewrite here,
\be
\deltak\left(\sqrt{-h}h^{ij}\right)=-8i\sqrt{-h}\left(h_{+}^{ik}h_{+}^{jl}\partial_{k}\brtheta\kappa_{l}+h_{-}^{ik}h_{-}^{jl}
\partial_{k}\brthetap\kappap_{l}\right)\,.
\label{kh2}
\ee
For consistency with 
\be
\delta(\sqrt{-h}h^{ij})=-\sqrt{-h}\left(h^{ik}h^{jl}-\half h^{kl}h^{ij}\right)\delta h_{kl}\,,
\ee
 the variation~\eqref{kh2} is, from \eqref{hpmprop}, \eqref{ehprop2},  symmetric and traceless,
\be
\ba{ll}
\deltak(\sqrt{-h}h^{ij})=\deltak(\sqrt{-h}h^{ji})\,,\quad&\quad h_{ij}\deltak(\sqrt{-h}h^{ji})=0\,.
\ea
\ee
With \eqref{kzeta} and \eqref{kh2} assumed, the variation of the Lagrangian spelled  in \eqref{deltaL2} simplifies, through some intermediate step~(\ref{AdeltaL3}), to
\be
\delta\cL_{\scriptscriptstyle{\rm{superstring}}}+\partial_{i}\left[\epsilon^{ij}\left(\Pi_{j}{}^{M}+\cA_{j}^{M}\right)\delta X_{M}\right]
=\epsilon^{ij}\Big[
\delta\cA_{iM}+i\left(h_{+i}{}^{k}h_{+}^{lm}\partial_{k}\brtheta\kappa_{m}
+h_{-i}{}^{k}h_{-}^{lm}\partial_{k}\brthetap\kappap_{m}
\right)\SDPi_{lM}
\Big]\SDPi_{j}^{M}\,.
\label{deltaL3}
\ee
The vanishing of the right hand side  of the above equality then should   fix  the kappa-symmetry transformation of the coordinate gauge symmetry potential. Yet, since the potential  is constrained  to satisfy $\cA_{i}^{M}\partial_{M}=0$ and $\cA_{i}^{M}\cA_{jM}=0$~\eqref{secconA},   it does not take the naive form one might be tempted to put:
\be
\delta\cA_{iM}\neq -i\left(h_{+i}{}^{k}h_{+}^{lm}\partial_{k}\brtheta\kappa_{m}
+h_{-i}{}^{k}h_{-}^{lm}\partial_{k}\brthetap\kappap_{m}
\right)\SDPi_{lM}\,.
\ee
Instead,  we must ``double'' this and project $\SDPi_{lM}$ to the coordinate gauge symmetry value,
\be
\deltak\cA_{iM}=-2i\left(h_{+i}{}^{k}h_{+}^{lm}\partial_{k}\brtheta\kappa_{m}
+h_{-i}{}^{k}h_{-}^{lm}\partial_{k}\brthetap\kappap_{m}
\right){\bf{\Big[}}\SDPi_{lM}{\bf{\Big]}}_{\scriptscriptstyle\rm{\bf projected}}\,,
\label{kappacA}
\ee
such that
\be
{\bf{\Big[}}\SDPi_{l}^{M}{\bf{\Big]}}_{\scriptscriptstyle\rm{\bf projected}}\times\partial_{M}=0\,.
\ee
Concretely,  the off-block diagonal form of the $\Ott$ invariant metric, $\cJ_{MN}$~\eqref{defcJ} naturally decomposes  all the doubled variables into two parts, such as 
\be
\ba{llll}
X^{M}=\left(\widetilde{X}_{\mu}\,,X^{\nu}\right)\,,\quad&\quad
\partial_{M}=\left(\widetilde{\partial}^{\mu}\,,\partial_{\nu}\right)\,,\quad&\quad
\Pi_{i}^{M}=\left(\widetilde{\Pi}_{i\mu}\,,\Pi_{i}^{\nu}\right)\,,\quad&\quad
\SDPi_{i}^{M}=\left(\widetilde{\SDPi}_{i\mu}\,,\SDPi_{i}^{\nu}\right)\,.
\ea
\label{decomp}
\ee
Without loss of generality up to $\Ott$ rotations, if we choose  the section by
\be
\ba{ll}
\partial_{M}=\left(0\,,\partial_{\nu}\right)\,,\quad&\quad\cA_{iM}\equiv\big(0\,,A_{i\mu}\big)\,,
\ea
\label{sectionchosen2}
\ee
we get
\be
{\bf{\Big[}}\SDPi_{lM}{\bf{\Big]}}_{\scriptscriptstyle\rm{\bf projected}}\equiv
\left(0\,,\,\widetilde{\SDPi}_{i\mu}\right)\,,
\ee
and thus,  the  kappa-symmetry transformation of the coordinate gauge symmetry potential~\eqref{kappacA} reads explicitly,
\be
\deltak A_{i\mu}=-2i\left(h_{+i}{}^{k}h_{+}^{lm}\partial_{k}\brtheta\kappa_{m}
+h_{-i}{}^{k}h_{-}^{lm}\partial_{k}\brthetap\kappap_{m}
\right)\widetilde{\SDPi}_{l\mu}\,.
\ee
After all, under the kappa-symmetry, the Lagrangian transforms to the total derivative,
\be
\deltak\cL_{\scriptscriptstyle{\rm{superstring}}}=-\partial_{i}\left[\epsilon^{ij}\left(\Pi_{j}{}^{M}+\cA_{j}^{M}\right)\deltak X_{M}\right]\,.
\ee
~\\

\section{Reductions to type IIA, IIB or IIC}
One of the characteristics in our construction of the superstring  action~\eqref{Action} -- as a counterpart to the maximally supersymmetric DFT~\cite{Jeon:2012hp} -- is the usage of not the conventional Riemannian variables,  $\{g_{\mu\nu},B_{\mu\nu},\phi\}$,  but the $\Ott$ covariant genuine DFT variables: in particular,~the DFT-vielbeins. They   represent  the doubled-yet-gauged spacetime  NS-NS background  on which the Green-Schwarz superstring propagates. As long as their  defining algebraic relations~\eqref{defV} are satisfied, our   superstring  action, as well as the target-spacetime  supersymmetric DFT, all work autonomically  without resorting to the Riemannian geometry or parametrization. The connection to the conventional Riemannian formulations, such as supergravities and the original Green-Schwarz superstring action, may follow if we solve the defining relations   in terms of  zehnbeins and $B$-field. Yet, there exists a class of configurations which do not admit such a Riemannian parametrization even locally at all~\cite{Lee:2013hma,Ko:2015rha} (\emph{c.f.}~\cite{Garcia-Fernandez:2013gja}).

Hereafter, for concreteness, yet without loss of generality, we fix the section as \eqref{sectionchosen}:
\be
\ba{ll}
\partial_{M}=\left(\tpartial^{\mu}\,,\partial_{\nu}\right)\equiv\left(0\,,\partial_{\nu}\right)\,,\quad&\quad\cA_{iM}\equiv\big(0\,,A_{i\mu}\big)\,.
\ea
\label{sectionchosen3}
\ee
~\\

\subsection{Type IIA or IIB\,: Riemannian backgrounds}
The DFT-vielbeins, $V_{Mp}$ and $\brV_{M\brp}$, are  ${20\times 10}$ matrices.  If their first half $10\times10$ square  blocks are non-degenerate, we may parametrize them as $\frac{1}{\sqrt{2}}(e^{-1})_{p}{}^{\mu}$ and $\frac{1}{\sqrt{2}}(\bre^{-1})_{\brp}{}^{\mu}$ respectively with some invertible matrices, $e_{\mu}{}^{p}$ and $\bre_{\mu}{}^{\brp}$. Then, the defining relations of the DFT-vielbeins~\eqref{defV} determine the remaining ${10\times10}$ blocks with one common free skew-symmetric field, $B_{\mu\nu}=-B_{\nu\mu}$~\cite{Jeon:2011cn,Jeon:2012kd},
\be
\ba{ll}
V_{Mp}=\textstyle{\frac{1}{\sqrt{2}}}{{\left(\ba{c} (e^{-1})_{p}{}^{\mu}\\(B+e)_{\nu p}\ea\right)}}\,,\quad
&\quad\brV_{M{\brp}}=\textstyle{\frac{1}{\sqrt{2}}}\left(\ba{c} (\bre^{-1})_{\brp}{}^{\mu}\\(B+\bre)_{\nu{\brp}}\ea\right)\,,
\ea
\label{Vform1}
\ee
while   $e_{\mu}{}^{p}$ and $\bre_{\nu}{}^{{\brp}}$ must meet  
\be
e_{\mu}{}^{p}e_{\nu}{}^{q}\eta_{pq}=-\bre_{\mu}{}^{{\brp}}\bre_{\nu}{}^{\brq}\breta_{\brp\brq}\,.
\label{ebre}
\ee
In \eqref{Vform1}, we  set, as usual, $B_{\mu p}=B_{\mu\nu}(e^{-1})_{p}{}^{\nu}$,  $B_{\mu\brp}=B_{\mu\nu}(\bre^{-1})_{{\brp}}{}^{\nu}$ and   $e_{\nu p}=e_{\nu}{}^{q}\eta_{qp}$, $\bre_{\nu\brp}=\bre_{\nu}{}^{\brq}\breta_{\brq\brp}$, \textit{etc}.  Of course, with respect to the choice of the section~\eqref{sectionchosen3}, $e_{\mu}{}^{p}$, $e_{\nu}{}^{\brp}$ and $B_{\mu\nu}$ are identified as a pair of zehnbeins corresponding to the common Riemannian metric, $g_{\mu\nu}$, and the NS-NS $B$-field. It follows that the DFT-metric, or  ``the generalized metric'',  is of the well-known form,
\be
\cH_{MN}=V_{Mp}V_{N}{}^{p}-\brV_{M\brp}\brV_{N}{}^{\brp}=
P_{MN}-\brP_{MN}=\left(\ba{cc}{{g^{-1}}}&{{-g^{-1}B}}\\{{Bg^{-1}}}&{{\,\,g-Bg^{-1}B}}\ea\right)\,.
\label{RiemannianGM}
\ee  
This is the most general parametrization of the DFT-metric satisfying the defining property~\eqref{defDFTmetric}, \textit{i.e.~}`a symmetric  $\ODD$ element',  \textit{if} the upper left $D\times D$ block is invertible. 

The existence of the pair of zehnbeins   reflects the very fact that the local Lorentz symmetry in DFT is  twofold,\footnote{The fact that the spin group is twofold  can lead to a phenomenological  prediction  to the Standard Model: the quarks and the leptons may belong to the distinct  spin groups~\cite{Choi:2015bga}.} \textit{i.e.~}${\Spint\times\oSpint}$.    It follows that  $(e^{-1}\bre)_{p}{}^{\brp}$ is a Lorentz rotation, 
 \be
(e^{-1}\bre)_{p}{}^{\brp}(e^{-1}\bre)_{q}{}^{\brq}\breta_{\brp\brq}=-\eta_{pq}\,,
\ee
and in particular,  
\be
\det(e^{-1}\bre)=\pm 1\,.
\label{detee}
\ee

Now, assuming the Riemannian parametrization of \eqref{Vform1}, from \eqref{usefulYZ}, our master Lagrangian~\eqref{Lagrangian} reduces to
\be
\ba{lll}
\cL_{\scriptscriptstyle{\rm{IIA/IIB}}}&=&-\half\sqrt{-h}h^{ij}\left[
\mr{\Pi}_{i}^{\mu}\mr{\Pi}_{j}^{\nu}g_{\mu\nu}
+\left(\mr{\widetilde{\Pi}}_{i\mu}-A_{i\mu}
+\mr{\Pi}_{i}^{\lambda}B_{\lambda\mu}\right)
\left(\mr{\widetilde{\Pi}}_{j\nu}-A_{j\nu}
+\mr{\Pi}_{j}^{\rho}B_{\rho\nu}\right)g^{\mu\nu}
\right]\\
{}&{}&-\epsilon^{ij}\left(
\mr{\Pi}_{i}^{\mu}A_{j\mu}-i\partial_{i}X^{\mu}\widetilde{\Sigma}_{j\mu}-i\partial_{i}\widetilde{X}_{\mu}\Sigma_{j}^{\mu}\right)\\
{}&=&-\sqrt{-h}h^{ij}
\mr{\Pi}_{i}^{\mu}\mr{\Pi}_{j}^{\nu}g_{\mu\nu}
+2i\epsilon^{ij}\partial_{i}X^{\mu}\widetilde{\Sigma}_{j\mu}
+\epsilon^{ij}\Sigma_{i}^{\mu}\widetilde{\Sigma}_{j\mu}
+\epsilon^{ij}\mr{\Pi}_{i}^{\mu}\mr{\Pi}_{j}^{\nu}B_{\mu\nu}
+\epsilon^{ij}\partial_{i}\widetilde{X}_{\mu}\partial_{j}X^{\mu}
\\
{}&{}&-\half\sqrt{-h}h^{ij}\left(\mr{\widetilde{\Pi}}_{i\mu}
+\mr{\Pi}_{i}^{\lambda}B_{\lambda\mu}
+\frac{\epsilon_{i}{}^{k}}{\sqrt{-h}}\mr{\Pi}_{k}^{\lambda}g_{\lambda\mu}
-A_{i\mu}\right)
\left(\mr{\widetilde{\Pi}}_{j\nu}
+\mr{\Pi}_{j}^{\rho}B_{\rho\nu}
+\frac{\epsilon_{j}{}^{l}}{\sqrt{-h}}\mr{\Pi}_{l}^{\rho}g_{\rho\nu}-A_{j\nu}\right)g^{\mu\nu}\,,
\ea
\label{typeIIAB}
\ee
where we put, like \eqref{decomp},
\be
\Sigma_{i}^{M}=\left(\widetilde{\Sigma}_{i\mu}\,,\,\Sigma_{i}{}^{\nu}\right)\,,
\ee
and we set without the coordinate gauge symmetry potential, 
\be
\ba{lll}
\mr{\Pi}_{i}^{M}:=\partial_{i}X^{M}-i\Sigma_{i}^{M}=\left(\mr{\widetilde{\Pi}}_{i\mu}\,,\,\mr{\Pi}_{i}^{\nu}\right)\,,\quad&\quad
\mr{\widetilde{\Pi}}_{i\mu}=\partial_{i}\widetilde{X}_{\mu}-i\widetilde{\Sigma}_{i\mu}\,,\quad&\quad
\mr{\Pi}_{i}^{\mu}=\partial_{i}X^{\mu}-i\Sigma_{i}^{\mu}\,.
\ea
\ee
The on-shell value of the coordinate gauge symmetry potential is,  from  the last line of \eqref{typeIIAB} which is a `perfect square' of the potential,
\be
A_{i\mu}\equiv\mr{\widetilde{\Pi}}_{i\mu}
+\mr{\Pi}_{i}^{\lambda}B_{\lambda\mu}
+\textstyle{\frac{\epsilon_{i}{}^{k}}{\sqrt{-h}}}\mr{\Pi}_{k}^{\lambda}g_{\lambda\mu}\,.
\label{onshellA}
\ee
Therefore,  after the auxiliary   potential being integrated out, our action reduces to
\be
\cS_{\scriptscriptstyle{\rm{IIA/IIB}}}={\textstyle{\frac{1}{2\pi\alpha^{\prime}}}}{\dis{\int}}\rd^{2}\sigma~
-\half\sqrt{-h}h^{ij}
\mr{\Pi}_{i}^{\mu}\mr{\Pi}_{j}^{\nu}g_{\mu\nu}
+\epsilon^{ij}\left(i\partial_{i}X^{\mu}
+\half\Sigma_{i}^{\mu}\right)\widetilde{\Sigma}_{j\mu}
+\half\epsilon^{ij}\mr{\Pi}_{i}^{\mu}
\mr{\Pi}_{j}^{\nu}B_{\mu\nu}
+\half\epsilon^{ij}\partial_{i}
\widetilde{X}_{\mu}\partial_{j}X^{\mu}
\,,
\label{ActionIIAB}
\ee
where  the standard string tension, $(2\pi\alphap)^{-1}$, is restored.
The last term, as total derivative,  is the topological term introduced in \cite{Giveon:1991jj} and \cite{Hull:2006va}. 

In order to compare with the original Green-Schwarz action, \textit{i)} we perform a $\oPint$ rotation  to let 
\be
e_{\mu}{}^{p}\equiv\bre_{\mu}{}^{\brp}\,,
\label{eequivbre}
\ee 
\textit{ii)}  truncate the twofold Lorentz symmetry to the diagonal subgroup,
\be
\ba{lll}
\Spint\times\oSpint&\quad\Longrightarrow\quad&
\Spint_{{\rm{Diagonal}}}\,,
\ea
\ee
and \textit{iii)} do not distinguish the  unbarred and barred spin group  indices anymore: in particular, we may identify
\be
\ba{lll}
\eta_{pq}\equiv-\breta_{\brp\brq}\,,\quad&\quad\brgamma^{\brp}\equiv\gammae\gamma^{p}\,,\quad&\quad\brgammae\equiv-\gammae\,.
\ea
\ee 
Then depending on the sign value of $\det(e^{-1}\bre)$ prior to the $\oPint$ rotation for \eqref{eequivbre}, the conventional  classification of  type IIA and type  IIB  can be recovered~\cite{Jeon:2012hp},
\be
\ba{lll}
\mbox{type~IIA}&\mbox{for}&
\det(e^{-1}\bre)=+1\,,\\
\mbox{type~IIB}&\mbox{for}&
\det(e^{-1}\bre)=-1\,.
\ea
\ee
Essentially, when $\det(e^{-1}\bre)=+1$ we can ensure the identification~\eqref{eequivbre} using $\oSpint$ group without flipping the chirality of $\brthetap$, but when $\det(e^{-1}\bre)=-1$ we have to use a chirality flipping $\oPint$ rotation.

In this way, setting ${B_{\mu\nu}=0}$,  up to the world-sheet topological term and  constant rescaling of the fermions, $\theta,\brtheta\rightarrow \sqrt[4]{2}\,\theta,\sqrt[4]{2}\,\brthetap$, the reduced  action~\eqref{ActionIIAB} can be identified as the original undoubled  Green-Schwarz superstring action.  \\

{\textit{\large{Self-duality over the entire doubled-yet-gauged spacetime.}} \\
 Since $\mr{\Pi}_{i}^{\mu}=\Pi_{i}^{\mu}$, \eqref{onshellA} is equivalent to 
\be
g^{\mu\nu}\widetilde{\Pi}_{i\nu}-(g^{-1}B)^{\mu}{}_{\nu}\Pi_{i}^{\nu}+\textstyle{\frac{\epsilon_{i}{}^{j}}{\sqrt{-h}}}\Pi_{j}^{\mu}=0\,.
\label{forSD0}
\ee
This gives,  contracting with $B_{\lambda\mu}$,
\be
(Bg^{-1})_{\lambda}{}^{\nu}\widetilde{\Pi}_{i\nu}
-(Bg^{-1}B)_{\lambda\nu}\Pi_{i}^{\nu}
+\textstyle{\frac{\epsilon_{i}{}^{j}}{\sqrt{-h}}}B_{\lambda\mu}\Pi_{j}^{\mu}=0\,,
\label{forSD1}
\ee
and further separately, contracting with $g_{\lambda\mu}\frac{\epsilon_{k}{}^{i}}{\sqrt{-h}}$,
\be
\textstyle{\frac{\epsilon_{i}{}^{j}}{\sqrt{-h}}}\left(\widetilde{\Pi}_{j\lambda}-B_{\lambda\nu}\Pi_{j}^{\nu}\right)+g_{\lambda\nu}\Pi_{i}^{\nu}=0\,.
\label{forSD2}
\ee
Adding (\ref{forSD1}) and (\ref{forSD2}), we obtain
\be
(Bg^{-1})_{\lambda}{}^{\nu}\widetilde{\Pi}_{i\nu}+
(g-Bg^{-1}B)_{\lambda\nu}\Pi_{i}^{\nu}+\textstyle{\frac{\epsilon_{i}{}^{j}}{\sqrt{-h}}}\widetilde{\Pi}_{j\lambda}=0\,.
\label{forSD3}
\ee
Then, as in the case with the bosonic string action of \cite{Lee:2013hma},  \eqref{forSD0} and \eqref{forSD3} imply that the full set of the self-duality relations hold over the entire doubled-yet-gauged spacetime coordinate directions -- although the coordinate gauge symmetry is a constrained field -- when the NS-NS background is Riemannian,
\be
\SDPi_{iM}=\Pi_{iM}+\textstyle{\frac{\epsilon_{i}{}^{j}}{\sqrt{-h}}}\cH_{M}{}^{N}\Pi_{jN}=\left(\SDPi_{i}^{\mu}\,,\,\widetilde{\SDPi}_{i\nu}\right)=0\,.
\label{SD2}
\ee
In the generic cases, \textit{i.e.~}not necessarily Riemannian, the equation of motion of the coordinate gauge symmetry potential gives \textit{a priori} only the half of the self-duality relations, from \eqref{deltaL},
\be
\ba{lll}
\epsilon^{ij}\delta\cA_{iM}\SDPi_{j}^{M}=0
\quad&\quad\Longrightarrow\quad&\quad
\SDPi_{i}^{\mu}=0\,.
\ea
\ee
Then the above result~\eqref{SD2} tells us that when the NS-NS background admits Riemannian interpretation, the other half of the self-duality relations is automatically satisfied, $\widetilde{\SDPi}_{i\nu}=0$. It is useful to note that, contracting  with the DFT-vielbeins, the self-duality~\eqref{SD2} decomposes into 
\be
\ba{ll}
h_{+}^{ij}\Pi_{jp}=0\,,\quad&\quad
h_{-}^{ij}\Pi_{j\brp}=0\,.
\ea
\label{SDg}
\ee
The self-dual part of $\Pi_{i}^{M}$ satisfies, from \eqref{ehprop1}, \eqref{ehprop2},
\be
\ba{ll}
\SDPi_{iM}=\textstyle{\frac{\epsilon_{i}{}^{j}}{\sqrt{-h}}}\cH_{M}{}^{N}\SDPi_{jN}\,,\quad&\quad
\epsilon^{ij}\SDPi_{jM}=\epsilon^{ij}\Pi_{jM}+\sqrt{-h}h^{ij}\cH_{M}{}^{N}\Pi_{jN}\,.
\ea
\ee
~\\


\subsection{Type IIC\,:  non-Riemannian and non-relativistic backgrounds}
While the non-Riemannian NS-NS background was first noted in  \cite{Lee:2013hma} and subsequently shown in \cite{Ko:2015rha} to lead to the Gomis-Ooguri non-relativistic bosonic string~\cite{Gomis:2000bd}, until now  there is no systematic classification of it.   Decomposing the DFT-vielbeins  in terms of $10\times 10$ square matrices, such as   $V_{Mp}=(V^{\mu}{}_{p},\widetilde{V}_{\nu p})$ and  $\brV_{M\brp}=(\brV^{\mu}{}_{\brp},\widetilde{\brV}_{\nu\brp})$, the   defining relations of them~(\ref{defV}), especially  the last one, imply
\be
V^{\mu}{}_{p}V^{\nu}{}_{q}\eta^{pq}=-\brV^{\mu}{}_{\brp}\brV^{\nu}{}_{\brq}\breta^{\brp\brq}\,.
\ee 
This shows that  $V^{\mu}{}_{p}$ is invertible if and only if  $\brV^{\mu}{}_{\brp}$ is so.  

In this subsection, we focus on the   non-Riemannian background  for  the Gomis-Ooguri non-relativistic string.  For this,   we  need to decompose  $\Ott$ into $\mathbf{O}(2,2)\times\mathbf{O}(8,8)$,  such that  the doubled coordinates decompose as $x^{M}=(x^{\hat{M}}, x^{M^{\prime}})$, where including the time directions we set $x^{\hat{M}}=(\tx_{\hat{\mu}},x^{\hat{\nu}})=(\tilde{t},\tilde{x}_{1},t,x^{1})$,  while $x^{M^{\prime}}$ denotes the remaining $16$ spatial part of the doubled-yet-gauged coordinates.  With respect to the choice of the section~\eqref{sectionchosen3}, the DFT-metric for the non-Riemannian NS-NS background reads~\cite{Lee:2013hma}
\be
\cH_{MN}=\left(\ba{cccc}0& 0&\hepsilon^{\hlambda}{}_{\hnu}&0\\
0&\delta^{\lambda^{\prime}\rho^{\prime}}&0&0\\
-\hepsilon_{\hmu}{}^{\hrho}&0&{\heta}_{\hmu\hnu}&0\\
0&0&0&\delta_{\mu^{\prime}\nu^{\prime}}\ea\right)\,,
\label{HformC}
\ee
where  we set  a two-dimensional flat Minkowskian metric, $\hat{\eta}_{\hmu\hnu}=\mbox{diag}(-+)$, and a skew-symmetric Levi-Civita symbol, $\hepsilon_{\hmu\hnu}$, with $\hepsilon^{\hlambda}{}_{\hnu}=\heta{}^{\hlambda\hmu}\hepsilon_{\hmu\hnu}$, $\hepsilon_{\hmu}{}^{\hrho}=\hepsilon_{\hmu\hnu}\heta{}^{\hnu\hrho}$, \textit{etc.} 
The upper left $2\times2$ block  is  vanishing completely  and hence clearly non-Riemannian in nature,  in contrast to the Riemannian generalized metric~\eqref{RiemannianGM}.

The corresponding DFT-vielbeins are essentially,
\be
\ba{ll}
\hat{V}_{\hat{M}\hat{p}}=\textstyle{\frac{1}{\sqrt{2}}}{{\left(\ba{rr} 
1&-1\\
-1&1\\
-1&0\\
0&1
\ea\right)}}\,,\quad
&\quad\hat{\brV}_{\hat{M}\hat{\brp}}=\textstyle{\frac{1}{\sqrt{2}}}\left(\ba{rr}
1&1\\
1&1\\
1&0\\
0&-1
\ea\right)\,.
\ea
\label{VformC}
\ee
As $4\times 2$ matrices, these represent  the genuinely non-Riemannian `hatted' part of the full DFT-vielbeins. The  remaining $16$ doubled-yet-gauged `primed' coordinates  is  flat Riemannian:  assigned the  flat Euclidean   kronecker-delta symbol, $\delta_{\mu^{\prime}\nu^{\prime}}$, as the spacetime metric with constant $B$-field.

The  master Lagrangian~\eqref{Lagrangian} reduces, upon the non-Riemannian NS-NS background, to 
\be
\ba{lll}
\cL_{\scriptscriptstyle{\rm{IIC}}}
&=&-\sqrt{-h}h^{ij}
\mr{\Pi}_{i}^{\mu}\mr{\Pi}_{j}^{\nu}\mr{g}_{\mu\nu}
+\epsilon^{ij}(2i\partial_{i}X^{\mu}+\Sigma_{i}^{\mu})\widetilde{\Sigma}_{j\mu}
+\epsilon^{ij}\partial_{i}\widetilde{X}_{\mu}\partial_{j}X^{\mu}
+\epsilon^{ij}
\mr{\Pi}_{i}^{\mu^{\prime}}\mr{\Pi}_{j}^{\nu^{\prime}}
B_{\mu^{\prime}\nu^{\prime}}
\\
{}&{}&+\sqrt{-h}\left(\cA_{i\hmu}-\mr{\widetilde{\Pi}}_{i\hmu}\right)
\left(h^{ij}\hepsilon^{\hmu}{}_{\hnu}\mr{\Pi}_{j}^{\hnu}+\frac{\epsilon^{ij}}{\sqrt{-h}}\mr{\Pi}_{j}^{\hmu}\right)\\
{}&{}&-\half\sqrt{-h}\,\left|\!\left|\,\mr{\widetilde{\Pi}}_{i\mu^{\prime}}
+\mr{\Pi}_{i}^{\lambda^{\prime}}B_{\lambda^{\prime}\mu^{\prime}}
+\frac{\epsilon_{i}{}^{k}}{\sqrt{-h}}\mr{\Pi}_{k}^{\lambda^{\prime}}\mr{g}_{\lambda^{\prime}\mu^{\prime}}
-A_{i\mu^{\prime}}\,\right|\!\right|^{2}\,,
\ea
\ee
where we set a  ten-dimensional target-spacetime  constant metric,
\be
\mr{g}_{\mu\nu}:=\left(\half\hat{\eta}_{\hmu\hnu}\,,\,\delta_{\mu^{\prime}\nu^{\prime}}\right)\,.
\ee
The resulting superstring action is then,
\be
\cS_{\scriptscriptstyle{\rm{IIC}}}={\textstyle{\frac{1}{2\pi\alpha^{\prime}}}}{\dis{\int}}\rd^{2}\sigma~
-\half\sqrt{-h}h^{ij}
\mr{\Pi}_{i}^{\mu}\mr{\Pi}_{j}^{\nu}\mr{g}_{\mu\nu}
+\epsilon^{ij}\left(i\partial_{i}X^{\mu}
+\half\Sigma_{i}^{\mu}\right)\widetilde{\Sigma}_{j\mu}
+\half\epsilon^{ij}\partial_{i}
\widetilde{X}_{\mu}\partial_{j}X^{\mu}+\half\epsilon^{ij}\mr{\Pi}_{i}^{\mu^{\prime}}
\mr{\Pi}_{j}^{\nu^{\prime}}B_{\mu^{\prime}\nu^{\prime}}\,,
\label{ActionIIC}
\ee
and is 
subject to the  chirality condition  for the   hatted  untilde  directions, $x^{\hmu}=(t,x^{1})\,$:
\be
\mr{\Pi}_{i}^{\hmu}+
\textstyle{\frac{\epsilon_{i}{}^{j}}{\sqrt{-h}}}\hepsilon^{\hmu}{}_{\hnu}\mr{\Pi}_{j}^{\hnu}= 0\,.
\label{nonRSD}
\ee
This is the action for the  Green-Schwarz superstring  on the non-Riemannian background which supersymmetrizes the  Gomis-Ooguri non-relativistic string.


\section{Discussion} 
In this work, we have constructed a world-sheet action for  Green-Schwarz superstring  which propagates   on  doubled-yet-gauged  spacetime.   For an arbitrarily curved NS-NS background, the action possesses manifest $\mathbf{O}(10,10)$ T-duality, $\mathbf{Spin}(1,9)\times\mathbf{Spin}(9,1)$ global Lorentz symmetry, coordinate gauge symmetry,  target-spacetime doubled-yet-gauged diffeomorphisms,  world-sheet diffeomorphisms and Weyl symmetry. Restricted to flat backgrounds of constant DFT-vielbeins, the action is further invariant  under     maximal  spacetime global supersymmetry  and also under local fermionic   kappa-symmetry.  After the   auxiliary  coordinate gauge symmetry potential being integrated out,  the action can  consistently reduce to the undoubled  original Green-Schwarz action upon a Riemannian  background. Thanks to the twofold spin groups, the action is unique: 
the two  fermions, $\theta^{\alpha}$ and $\thetap{}^{\bralpha}$,  are Majora-Weyl spinors  for $\Spint$ and  $\oSpint$ respectively. It is then  specific choices of the NS-NS backgrounds that distinguish  Riemannian   IIA, IIB and  non-Riemannian IIC.   Upon the Riemmanian  IIA/IIB  backgrounds, the Euler-Lagrangian equation of the coordinate gauge symmetry potential implies the self-duality over the entire doubled-yet-gauged spacetime.

It will be of interest to couple our action to  the   $\Spint\times\oSpint$ bi-spinorial R-R sector~\cite{Jeon:2012kd}. Investigating the  supersymmetry, the Killing spinor equations of the maximally supersymmetric DFT~\cite{Jeon:2012hp} should appear naturally. The computations of the  one-loop beta function and  the partition function are worth while to perform: we expect to derive the  equations of motion of the maximally supersymmetric DFT~\cite{Jeon:2012hp}. Related to this,  we refer readers to earlier works~\cite{Berman:2007xn,Berman:2007yf,Tan:2014mba} on  bosonic doubled sigma models, along with \cite{Ko:2015rha} for the  matching of the fluctuation spectrum  between  DFT and  the bosonic world-sheet action~\eqref{bosonicaction} around the non-Riemannian background for the Gomis-Ooguri string.  
Promoting the global $\Spint\times\oSpint$ Lorentz symmetry to the local symmetry seems desirable. 
We leave  quantization  as for future work.


\section*{Acknowledgements}
 We wish to thank Kanghoon Lee, Charles M. Melby-Thompson and Ren\'e Meyer for  discussions. We also thank  an anonymous referee for suggesting  us to look for the Wess-Zumino term~(\ref{WZ}).     This work was  supported by the National Research Foundation of Korea with Grant Nos. 2015K1A3A1A21000302 and 2016R1D1A1B0101519. \\


\appendix

\section{Useful identities}

In addition to \eqref{Fierztd2}, the Fierz identity~\eqref{Fierztd} implies for arbitrary $\delta\theta$ and $\delta\thetap$,
\be
\ba{lll}
\epsilon^{ij}\partial_{i}(\brtheta\gamma^{p}\delta\theta)\brtheta\gamma_{p}\partial_{j}\theta&=&
\textstyle{\frac{1}{3}}
\partial_{i}\left(\epsilon^{ij}\brtheta\gamma^{p}\delta\theta\brtheta
\gamma_{p}\partial_{j}\theta\right)+
\textstyle{\frac{2}{3}}
\epsilon^{ij}\brtheta\gamma^{p}(\partial_{i}\delta\theta)\brtheta\gamma_{p}\partial_{j}\theta\,,\\

\epsilon^{ij}\partial_{i}(\brthetap\brgamma^{\brp}\delta\thetap)\brthetap\brgamma_{\brp}
\partial_{j}\thetap&=&
\textstyle{\frac{1}{3}}
\partial_{i}\left(\epsilon^{ij}\brthetap\brgamma^{\brp}
\delta\thetap\brthetap\brgamma_{\brp}\partial_{j}\thetap\right)+
\textstyle{\frac{2}{3}}
\epsilon^{ij}\brthetap\brgamma^{\brp}(\partial_{i}\delta\thetap)\brthetap\brgamma_{\brp}\partial_{j}\thetap\,,
\ea
\label{AFierz1}
\ee
and
\be
\ba{lll}
\epsilon^{ij}\brtheta\gamma^{p}(\partial_{i}\delta\theta)\,\brtheta\gamma_{p}\partial_{j}\theta&=&
\partial_{i}\left(\epsilon^{ij}\brtheta\gamma^{p}\delta\theta\brtheta
\gamma_{p}\partial_{j}\theta\right)+
3\epsilon^{ij}\delta\brtheta\gamma^{p}\partial_{i}\theta\brtheta
\gamma_{p}\partial_{j}\theta\,,\\
\epsilon^{ij}\brthetap\brgamma^{\brp}(\partial_{i}\delta\thetap)\,\brthetap\brgamma_{\brp}\partial_{j}\thetap&=&
\partial_{i}\left(\epsilon^{ij}\brthetap\brgamma^{\brp}\delta\thetap\brthetap
\brgamma_{\brp}\partial_{j}\thetap\right)+
3\epsilon^{ij}\delta\brthetap\brgamma^{\brp}\partial_{i}\thetap\brthetap
\brgamma_{\brp}\partial_{j}\thetap\,.
\ea
\label{Afierz2}
\ee
It is worth while to note
\be
\ba{ll}
\epsilon^{ij}\epsilon^{kl}=h\left(h^{ik}h^{jl}-h^{jk}h^{il}\right)\,,\quad&\quad
\epsilon^{-1}_{ij}\epsilon^{-1}_{kl}=h^{-1}\left(h_{ik}h_{jl}-h_{jk}h_{il}\right)\,,\\
\epsilon^{i}{}_{j}\epsilon^{j}{}_{k}=\epsilon^{il}\epsilon^{jm}h_{lj}h_{mk}=-h\delta^{i}_{~k}\,,\quad&\quad
\epsilon^{-1}_{ij}=-h^{-1}h_{ik}h_{jl}\epsilon^{kl}=-h^{-1}\epsilon_{ij}\,,
\ea
\label{ehprop1}
\ee
\be
\ba{ll}
h^{i}_{\pm j}\textstyle{\frac{\epsilon^{jk}}{\sqrt{-h}}}=
h^{i}_{\pm j}(h^{jk}_{+}-h^{jk}_{-})=\pm h^{ik}_{\pm }\,,\quad&\quad 
\sqrt{-h}h^{ij}_{\pm}\epsilon^{-1}_{jk}=\pm h^{i}_{\pm k}\,,
\\
\textstyle{\frac{\epsilon^{ij}}{\sqrt{-h}}}h_{\pm j}{}^{k}=\pm h_{\pm}^{ik}\,,\quad&\quad
\sqrt{-h}\epsilon^{-1}_{ij}h^{jk}_{\pm}=\pm h_{\pm i}{}^{k}\,,\\
h^{ij}_{\pm}h^{kl}_{\pm}=h^{il}_{\pm}h^{kj}_{\pm}\,,\quad&\quad
h_{\pm}^{ik}h_{\pm}^{jl}
\left(\epsilon_{k}{}^{m}\delta_{l}^{~n}-\epsilon_{l}{}^{n}\delta_{k}^{~m}\right)=0\,,
\ea
\label{ehprop2}
\ee
\be
\ba{ll}
h_{\pm i}^{i}=1\,,\quad&\quad\det(h_{\pm}^{ij})=0\,,
\ea
\label{ehprop3}
\ee
and
\be
\ba{ll}
\gamma_{pq}\Pi^{p}_{i}\Pi^{q}_{j}=-\half\epsilon^{-1}_{ij}\epsilon^{kl}
\gamma_{pq}\Pi^{p}_{k}\Pi^{q}_{l}\,,\quad&\quad
\brgamma_{\brp\brq}\Pi^{\brp}_{i}\Pi^{\brq}_{j}=
-\half\epsilon^{-1}_{ij}\epsilon^{kl}\brgamma_{\brp\brq}
\Pi^{\brp}_{k}\Pi^{\brq}_{l}\,.
\ea
\label{gamma2e}
\ee

Substituting the ansatz~\eqref{deltatheta} into \eqref{deltaL}, the variation of the Lagrangian reduces to  
\be
\ba{lll}
\delta\cL_{\scriptscriptstyle{\rm{superstring}}}&=&-\half
\left[\delta\left(\sqrt{-h}h^{ij}\right)+4i\sqrt{-h}\left(\partial_{k}\brtheta\zeta^{(i}h_{+}^{j)k}+
\partial_{k}\brthetap\zetap{}^{(i}h_{-}^{j)k}\right)\right]
\Pi_{i}^{M}\Pi_{j}^{N}\cH_{MN}\\
{}&{}&-2i\sqrt{-h}\left(
\partial_{k}\brtheta\zeta^{(i}h_{+}^{j)k}-
\partial_{k}\brthetap\zetap{}^{(i}h_{-}^{j)k}\right)\Pi_{iM}\Pi_{j}^{M}
+\left(\sqrt{-h}h^{ij}\cH^{M}{}_{N}\Pi_{j}^{N}+\epsilon^{ij}\Pi_{j}^{M}\right)\delta\cA_{iM}\\
{}&{}&-2i\left(h_{-j}^{i}\partial_{i}\brtheta\gamma_{MN}\zeta^{j}+h_{+j}^{i}\partial_{i}
\brthetap\brgamma_{MN}\zetap{}^{j}\right)
\left(\epsilon^{kl}\Pi_{k}^{M}\Pi_{l}^{N}\right)
-\partial_{i}\left[\epsilon^{ij}\left(\Pi_{j}{}^{M}+\cA_{j}^{M}\right)\delta X_{M}\right]\\
{}&=&-\half
\left[\delta\left(\sqrt{-h}h^{ij}\right)+8i\sqrt{-h}\left(\partial_{k}\brtheta\zeta^{(i}h_{+}^{j)k}+
\partial_{k}\brthetap\zetap{}^{(i}h_{-}^{j)k}\right)\right]
\Pi_{i}^{M}\Pi_{j}^{N}\cH_{MN}\\
{}&{}&+\left(\sqrt{-h}h^{ij}\cH^{M}{}_{N}\Pi_{j}^{N}+\epsilon^{ij}\Pi_{j}^{M}\right)\Big[
\delta\cA_{iM}+2i\left(h_{+i}{}^{k}\partial_{k}\brtheta\zeta^{l}+h_{-i}{}^{k}\partial_{k}\brthetap\zetap^{l}\right)\Pi_{lM}\Big]\\
{}&{}&-2i\left(h_{-j}^{i}\partial_{i}\brtheta\gamma_{MN}\zeta^{j}
+h_{+j}^{i}
\partial_{i}\brthetap\brgamma_{MN}\zetap{}^{j}\right)
\left(\epsilon^{kl}\Pi_{k}^{M}\Pi_{l}^{N}\right)
-\partial_{i}\left[\epsilon^{ij}\left(\Pi_{j}{}^{M}+\cA_{j}^{M}\right)\delta X_{M}\right]\,.
\ea
\label{AdeltaL2}
\ee
This further becomes, with \eqref{kzeta} and \eqref{kh2},
\be
\ba{l}
\delta\cL_{\scriptscriptstyle{\rm{superstring}}}
+\partial_{i}\left[\epsilon^{ij}\left(\Pi_{j}{}^{M}+\cA_{j}^{M}\right)\delta X_{M}\right]\\
=\Big[\epsilon^{ij}
\delta\cA_{iM}+2i\sqrt{-h}\left(-h_{+}^{jk}h_{+}^{lm}\partial_{k}\brtheta\kappa_{m}
+h_{-}^{jk}h_{-}^{lm}\partial_{k}\brthetap\kappap_{m}
\right)\Pi_{lM}
\Big]\SDPi_{j}^{M}\\
=\Big[\epsilon^{ij}
\delta\cA_{iM}+i\sqrt{-h}\left(-h_{+}^{jk}h_{+}^{lm}\partial_{k}
\brtheta\kappa_{m}
+h_{-}^{jk}h_{-}^{lm}\partial_{k}\brthetap\kappap_{m}
\right)\SDPi_{lM}
\Big]\SDPi_{j}^{M}\\
=\epsilon^{ij}\Big[
\delta\cA_{iM}+i\left(h_{+i}{}^{k}h_{+}^{lm}\partial_{k}\brtheta\kappa_{m}
+h_{-i}{}^{k}h_{-}^{lm}\partial_{k}\brthetap\kappap_{m}
\right)\SDPi_{lM}
\Big]\SDPi_{j}^{M}\,,
\ea
\label{AdeltaL3}
\ee
where for the second equality we have used  identities, 
\be
\ba{ll}
h_{+}^{jk}h_{+}^{lm}\SDPi_{jM}\Pi_{l}^{M}=\half
h_{+}^{jk}h_{+}^{lm}\SDPi_{jM}\SDPi_{l}^{M}\,,\quad&\quad
h_{-}^{jk}h_{-}^{lm}\SDPi_{jM}\Pi_{l}^{M}=\half
h_{-}^{jk}h_{-}^{lm}\SDPi_{jM}\SDPi_{l}^{M}\,.
\ea
\ee

One useful relation to establish  the second equality in \eqref{typeIIAB}   is
\be
\ba{l}
-\half\sqrt{-h}h^{ij}(\tY_{i\mu}-A_{i\mu})(\tY_{j\nu}-A_{j\nu})g^{\mu\nu}-\epsilon^{ij}Z_{i}{}^{\mu}A_{j\mu}\\
=-\half\sqrt{-h}h^{ij}\left(\frac{\epsilon_{i}{}^{k}}{\sqrt{-h}}Z_{k}{}^{\lambda}g_{\lambda\mu}+\tY_{i\mu}-A_{i\mu}\right)\left(\frac{\epsilon_{j}{}^{l}}{\sqrt{-h}}Z_{l}{}^{\rho}g_{\rho\nu}+\tY_{j\nu}-A_{j\nu}\right)g^{\mu\nu}\\
\,\quad-\half\sqrt{-h}h^{ij}Z_{i}{}^{\mu}Z_{j}{}^{\nu}g_{\mu\nu}-\epsilon^{ij}Z_{i}{}^{\mu}\tY_{j\mu}\,.
\ea
\label{usefulYZ}
\ee
This identity  is true for arbitrary $\tY_{i\mu}$ and $Z_{j}{}^{\nu}$.


\end{document}